\documentclass[a4paper]{jpconf}

\usepackage{citesort,iopams}
\usepackage{graphicx}

\usepackage{sidecap}

\newcommand{\bd}{\begin{displaymath}}
\newcommand{\ed}{\end{displaymath}}
\newcommand{\be}{\begin{equation}}
\newcommand{\ee}{\end{equation}}
\newcommand{\ba}{\begin{eqnarray}}
\newcommand{\ea}{\end{eqnarray}}

\begin{document}

\title{Quantum Interference, Hidden Symmetries:\\
Theory and Experimental Facts}

\author{\'Angel S. Sanz}

\address{Department of Optics, Faculty of Physical Sciences,
Universidad Complutense de Madrid,\\
Pza.\ Ciencias 1, Ciudad Universitaria E-28040 Madrid, Spain}

\ead{a.s.sanz@fis.ucm.es}

\begin{abstract}
The concept of quantum superposition is reconsidered and discussed from the viewpoint
of Bohmian mechanics, the hydrodynamic formulation of quantum mechanics, in order to elucidate
some physical consequences that go beyond the simple mathematical idea of linearly combining vectors
in a Hilbert space.
Specifically, the discussion turns around the connection between symmetries characterizing
the wave function and the behavior in time displayed by the quantum flux when the latter is analyzed
in terms of streamlines (Bohmian trajectories).
This is illustrated with a series of analytical results and numerical simulations, which include
Young's two-slit experiment, counter-propagating wave packets, grating diffraction and quantum
carpets (e.g., Talbot carpets), and diffraction under confinement conditions.
From the analysis presented it follows that quantum paradoxes appear whenever symmetries related
to interference are neglected in the interpretation and understanding of the corresponding phenomena.
\end{abstract}


\section{Introduction}
\label{sec1}

\vspace{.25cm}

\begin{quote}
In making some experiments on the fringes of colours accompanying
shadows, I have found so simple and so demonstrative a proof of the general
law of the interference of two portions of light [\ldots].
The proposition on which I mean to insist at present, is simply this,
that fringes of colours are produced by the interference of two portions of
light; [\ldots] the assertion is proved by the experiments I am about to relate
[\ldots].
\end{quote}
With these remarkable words, Thomas Young started {\it Experiments and calculations
relative to physical optics} \cite{young:PTRSL:1804}, published by the end of 1804, although
presented to the Royal Society a year before, the 24th of November of 1803.
In this work, already considered a classic of the physics literature, Young provides
the first empirical evidence of the wave nature of light through one of its most
recognizable traits, namely the interference phenomenon.
This experiment overthrew 100 years of Newton's corpuscular conception of light\footnote{Newton published
his {\it Opticks} precisely in 1704, about a century earlier than Young's experimental finding.} and
strongly supported Huygens' wave view.
The latter would become the dominant trend until 1905.
Ironically, just a century after Young's work, Einstein went back to Newton's model, introducing the concept
of quantum of light or radiation ({\it Lichtquant}) to explain the photoelectric effect \cite{einstein:AnnPhys:1905-1}.

\begin{figure}[t]
\centering
  \includegraphics[width=12cm]{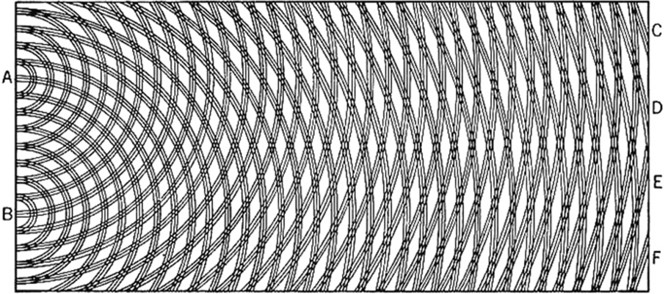}
  \caption{\label{fig1}
   Young's two-slit experiment \cite{young:PTRSL:1804}.
   Explanation using the Huygens construction with semicircular wavefronts generated at the source slits A and B.
   In the outgoing or diffracted waves, maxima are denoted with a higher accumulation of semicircles (three closely packed
   semicircles) and minima with absence of them; the alternation of maxima and minima gives an idea of the concept of
   wavelength.
   Far from the slits, on the opposite side of the figure, the wavefronts may coincide (overlap), giving rise to
   interference maxima; on the contrary, they may also arrive with half a wavelength of delay one another, generating
   minima (at the locations indicated with the letters C, D, E and F).}
\end{figure}

Essentially, what Young observed is the following.
Consider light coming from a single source (e.g., sunlight), passing through a hole.
If an obstacle with two nearly similar apertures is positioned in the way of the light,
it splits up.
Light then continues traveling until it reaches another distant screen (see Fig.~\ref{fig1}).
At this screen, instead of observing essentially two spots of light, there is a redistribution
of the light in the form of an alternating pattern of dark and bright spots or fringes.
This pattern cannot be in any way accounted for by means of the notion of light ray from geometric
optics, as happens with reflection, refraction or even mirage phenomena.
To explain the observed phenomenon, Young had to reject Newton's approach and consider Huygens'
his wavefront-based construction, as seen in Fig.~\ref{fig1}.
Thus, after passing through each opening, A or B, the emerging wavefronts increase their radii until they
start overlapping.
In some regions, the maxima of the wavefronts coincide and we observe a reinforcement of the light
intensity ({\it constructive interference}).
In other regions, maxima coincide with minima, thus
leading to a cancelation of light ({\it destructive interference}), which is precisely
what we have at the spots C, D, E and F in Fig.~\ref{fig1}.
In other words, light plus light may equal shadow.

Up to date, Young's experiment constitutes a powerful proof of the wave nature of light, which
has reinforced the concept of superposition in physics, with notorious applications in optics
and quantum mechanics.
In the later case, it has a special practical relevance.
Consider a set of solutions $|\Psi_i\rangle$, with $i= 1, 2, 3, \ldots$, all satisfying the time-dependent Schr\"odinger
equation,
\be
  i \hbar \frac{\partial |\Psi_i\rangle}{\partial t} = \hat{H} |\Psi_i \rangle .
 \label{eq1}
\ee
Additional solutions can be readily constructed by picking up different vectors from this set and
linearly combining them in various proportions, i.e.,
\be
 |\Psi\rangle = c_1|\Psi_1\rangle + c_2|\Psi_2\rangle + c_3|\Psi_3\rangle + \cdots
 \label{eq2}
\ee
In this linear combination, the $c_i$ contain information about both the contribution of each
$|\Psi_i\rangle$ to the superposition (through the weight $|c_i|^2$) and the mutual phase difference
between every two of these vectors [through the phase $(\tan)^{-1}({\rm Im}\{c_i c_j^*\}/{\rm Re}\{c_i c_j^*\})$,
with $i \ne j$].
The sum (\ref{eq2}) is a generalization of Young's empirical result very
convenient to solve complex quantum problems by means of known (eigen)solutions (the so-called
spectral methods \cite{Shizgal-bk:2015}, for instance, are based on this idea), and also to investigate the behavior
exhibited by quantum systems acted by slits, gratings or lattices (e.g., electron, neutron or molecular diffraction,
regardless of the mass and/or complexity of the probing particle \cite{arndt:NNanotech:2012,arndt:NaturePhys:2014}).

Apart from its practical orientation, the above idea also introduces a rather natural and deep question: has physical superposition
the same meaning as mathematical superposition?
In a real experiment one always obtains results that, after analysis, can be recast or modeled in terms of
sums like (\ref{eq2}).
However, the experiment itself is a whole that cannot be decomposed in the same way.
A simple illustration of this fact is the experiment performed in 1989 by Akira Tonomura and his
group at the Hytachi Central Research Laboratory,\footnote{It is worth noting that the tradition in
event-by-event experiments (i.e., particle-by-particle in the present context), produced with dim light, precedes quantum
mechanics itself, dating back to the experiments of Geoffrey Ingram Taylor in the first decade of the
XXth century \cite{taylor:ProcCambPhilSoc:1909}.
On the other hand, it is also to be highlighted the first realization of Young's two-slit experiment
with matter waves, specifically with electrons, by Claus J\"onsson \cite{jonsson:ZPhys:1961,jonsson:AJP:1974},
as well as the works in the same line of Giulio Pozzi and coworkers a few years later \cite{pozzi:AJP:1973},
who also produced the first experimental evidence of electron-by-electron interference patterns \cite{pozzi:AJP:1976}.}
consisting in reproducing Young's experiment
with the electrons expelled by an electron microscope \cite{tonomura:ajp:1989}.
In this renowned experiment, electrons are detected individually after having passed through a
magnetic equivalent to Young's two slits.
Initially, electrons reach the detector following a sort of random distribution.
However, as time proceeds and the number of electrons detected becomes larger, an interference
pattern starts emerging, where alternating dark and bright fringes can be readily identified.
Bright fringes denote regions of higher number of arrivals (i.e., high probability), while dark
regions indicate locations that electrons tend to avoid.
Accordingly, interference is only going to be observable if the sum (\ref{eq2}), for $i=2$, is taken
as a whole, even if mathematically it does not matter to proceed with separate waves \cite{sanz:JPA:2008}.

What is then special about having two slits open at the same time beyond the superposition principle?
To provide an answer to this question, we are going to appeal here to the notion of symmetry, in particular
to a kind of symmetry associated with the full wave function whenever it is split up, as in the case of Young's
experiment, that has nothing to do with the usual ones related to conserved quantities.
For instance, when matter waves are diffracted by (transmission or reflection) periodic gratings, as
mentioned above, Talbot-like patterns arise in the near field (Fresnel regime) \cite{sanz:AOP:2015}.
On the other hand, when dealing with scattering off periodic structures, it is known that diffraction
features in the far field (Fraunhofer regime) can be directly characterized from the periodic
properties of such structures \cite{sanz:SSR:2004}.
This is what happens, for example, in solid state physics, where diffraction features and energy band structures are
essentially obtained from symmetry properties associated with a single unit cell (Bravais lattice), that is, the unit
from which the whole of an ideal crystal solid is formed by repetition (translational invariance).
This constitutes the
physical basis for the periodic Born-von Karman boundary conditions and Bloch's theorem \cite{ashcroft-bk}.
Thus, these symmetries have to do precisely with the physical manifestation of the superposition (\ref{eq2}), which
avoids dealing with the system dynamics in terms of separated waves, but their combination has to be
considered as a whole, as mentioned above.

The discussion here thus turns around the physical implications of the symmetries displayed or
associated with coherent superpositions, i.e., wave functions that can be recast in the form of the sum (\ref{eq2}).
To this end, instead of using the typical language in terms of fields or operators, of common use in quantum mechanics,
the issue will be tackled with the aid of Bohmian mechanics \cite{bohm:PR:1952-1,bohm:PR:1952-2,holland-bk},
i.e., its hydrodynamic formulation \cite{sanz:EJP-arxiv:2017}.
In it, quantum systems are described as if they were
diffusing fluids in a configuration space that can be tracked by means of ensembles of streamlines or trajectories.
These ensembles of trajectories are characterized by two properties relevant to our discussion: (1) their
gathering and evolution in time is in compliance with the natural evolution of standard wave functions (i.e., they
do not arise from any sort of approximation), and (2) this evolution enhances the role of the quantum phase as
the main agent ruling quantum dynamics.
Accordingly, the work has been organized as follows.
The main elements of the Bohmian approach are introduced and discussed in Section~\ref{sec2}.
In Section~\ref{sec3} some general aspects on quantum superpositions and their analysis with Bohmian mechanics
are discussed before starting a discussion on numerical results obtained for different paradigmatic examples, including
Young's two-slit experiment, counter-propagating wave packets, grating diffraction and quantum carpets (e.g.,
Talbot carpets), and diffraction under confinement conditions.
The work concludes with a series of remarks summarized in Section~\ref{sec4}.


\section{About Bohm's quantum picture: A convenient working tool}
\label{sec2}

As in classical mechanics, there are different formulations equally legitimate
in quantum mechanics (e.g., Schr\"odiner's wave mechanics, Heisenberg's matrix formulation, Feynman's path
integral approach, etc.), each exploring or based on a different aspect of the quantum phenomenon or
process to be analyzed.
Using one or another will depend essentially on the simplicity with which we can treat the
corresponding problem \cite{sanz:EJP-arxiv:2017}.
One of such formulations is the quantum hydrodynamic approach, also known as Bohmian
mechanics.
In this formulation, apart from the wave function (or wave field), there is a set of associated auxiliary
trajectories that help us to understand how the system spreads throughout configuration space.
Initially, such trajectories were conceived as hidden variables \cite{bohm:PR:1952-1}, because
they allowed to monitor the evolution of the quantum system without perturbing its wave function.
However, here we are going to leave aside this ontological view (interesting, although out of
the scope of the current work) and focus
on a more pragmatic level, regarding the trajectories just as streamlines that help us to visualize
the time-evolution of the matter wave in space, in analogy to what happens in classical
fluid mechanics \cite{sanz:JPhysConfSer:2014}.
This view arises from a joint combination of three experimental facts
that the Bohmian description nicely gathers:
\begin{itemize}
 \item Quantum phenomena occur in real time.
  Hence the approach to be used in their description should also include
  this feature, rendering a view of the phenomenon analyzed as evolving in time.

 \item In an experiment, quantum observables are just statistical outcomes (something particularly
  evident in low-intensity experiments \cite{tonomura:ajp:1989,arndt:NNanotech:2012}), so the approach should
  also allow us to produce the experimental outcomes by means of a specific, non-ambiguous statistical
  procedure.

 \item Typically, in such experimental statistics, there is no time-correlation among different detections
  (detected events) \cite{pozzi:EJP:2013}, which means that the evolution of one event (a diffracted particle, for example)
  is unaware of the rest of events (the other diffracted particles, past and future, that contribute to
  the final diffraction pattern).
\end{itemize}
Simulators based on these three facts seem to be ideal candidates to explore the physics of quantum
systems beyond the bare density approach typically enabled by other quantum pictures.

Now, what are the essential ingredients of Bohmian mechanics?
In the case of Bohmian mechanics, the starting point, as formerly considered by Madelung \cite{madelung:ZPhys:1926}
or Bohm \cite{bohm:PR:1952-1,bohm:PR:1952-2}, is Schr\"odinger's picture, where the wave function accounting
for the state of the quantum system in the position representation is recast in the form of two real-valued fields instead
of a complex-valued one by virtue of the nonlinear polar transformation
\be
 \Psi({\bf r},t) = A({\bf r},t) e^{iS({\bf r},t)/\hbar} .
 \label{eq3}
\ee
From this relation, we find that the amplitude $A$ is directly connected to the probability density,
\be
 \rho = \Psi^* \Psi = A^2 ,
 \label{eq4}
\ee
while the phase field, defined as
\be
 S = \frac{\hbar}{2i}\ \ln \left( \frac{\Psi}{\Psi^*} \right) ,
 \label{eq5}
\ee
is going to be related to the quantum current density or quantum flux, ${\bf J}$, as seen below.
After substitution of the ansatz (\ref{eq3}) into the time-dependent Schr\"odinger equation,
\be
 i\hbar\ \frac{\partial \Psi}{\partial t} = -\frac{\hbar^2}{2m}\ \! \nabla^2 \Psi + V \Psi ,
 \label{eq6}
\ee
we obtain two coupled real-valued partial differential equations of motion ruling
the evolution in space and time of the fields (\ref{eq4}) and (\ref{eq5}):
\setlength\arraycolsep{2pt}
\begin{eqnarray}
 \frac{\partial \rho}{\partial t} & + &
  \nabla \! \cdot \! \left( \rho\ \! \frac{\nabla S}{m} \right) = 0 ,
 \label{eq7} \\
 \frac{\partial S}{\partial t} & + & \frac{(\nabla S)^2}{2m} + V + Q = 0 .
 \label{eq8}
\end{eqnarray}
Equation~(\ref{eq7}) is the usual continuity equation, which accounts for the conservation of the
probability, while Eq.~(\ref{eq8}) was regarded by Bohm as a quantum
Hamilton-Jacobi equation due to its formal similarity with its classical homolog, although the presence of
the $Q$ term, discussed below, sets a remarkable difference.
In Eq.~(\ref{eq8}), the phase field $S$ thus becomes a generalization of the classical mechanical action (or the
eikonal, in the context of optics), which was used by Bohm to postulate an associated momentum according to
an also generalized Jacobi's law,
\be
 {\bf p}_B = \nabla S .
 \label{eq9}
\ee
In classical mechanics, this law relates the generalized momentum in phase space with the surfaces
of constant action, while here it allows to define a momentum in the position representation (not to be confused with
the variable defining the momentum representation), which is perpendicular to surfaces of constant phase.
The quantum phase does play an important role in the dynamical behavior displayed by quantum systems, which
is typically disregarded because it is not an observable in strict sense (although its effects cannot be
neglected).

From Eq.~(\ref{eq9}) an equation of motion for trajectories or streamlines can be thus straightforwardly
obtained:
\be
  \dot{\bf r} = \frac{{\bf p}_B}{m} = \frac{\nabla S}{m}
   = \frac{1}{m}\ \! \frac{{\rm Re} \left\{ \Psi^* \hat{\bf p} \Psi \right\}}{|\Psi|^2} ,
  \label{eq10}
\ee
which is what Bohm did back in 1952, showing that the concept of trajectory, obtained after integration of
this equation, is not in contradiction at all with quantum mechanics.
Notice in the term after the third equality that $\hat{\bf p} = -i\hbar\nabla$ is the usual momentum operator
in the position representation, which means that the equation of motion can be understood as a sort of
expectation value for the momentum operator for the quantum state $\Psi$ (or, dividing by the mass $m$, for
an associated velocity operator).
Nevertheless, it is convenient to bear in mind that, although Bohm introduced the trajectories after postulating
the quantum Jacobi law, this is not a necessary condition, since it can be directly generated from usual
quantum mechanics through the standard definition of the quantum flux \cite{schiff-bk} by noting that
\be
 {\bf J} = \frac{1}{m}\ \! {\rm Re} \left\{ \Psi^* \hat{\bf p} \Psi \right\}
   = \frac{\nabla S}{m}\ \rho .
 \label{eq11}
\ee
This relation establishes a bridge between standard quantum tools and Bohmian mechanics.
If, according to the continuity equation, we assume the probability density is ``transported'' through configuration
space by means of a certain velocity field,\footnote{Transport phenomena are described by the diffusion equation.
In this regard, it is interesting to note the formal analogy between Schr\"odinger's equation and
the diffusion equation on the basis that they both are parabolic partial differential equations (see, for instance,
Ch.~12 in \cite{morse-bk-2}).
This was an aspect already noticed by F\"urth \cite{furth:ZPhys:1933} or, later on, also by Comisar \cite{comisar:PhysRev:1965},
identifying the quantity $i\hbar/2m$ with an imaginary diffusion constant.
However, the analogy breaks when we noticed that the diffusion equation accounts for a conservation law, while the same does not hold for
Schr\"odinger's equation, since the conserved quantity is the probability density and not the probability amplitude
or wave function.}
thus generating a current density, i.e., ${\bf J} = {\bf v} \rho$, then we readily find that such a velocity
field is proportional to Bohm's momentum, ${\bf v} = {\bf p}_B/m$, as can be seen from (\ref{eq11}).
The possibility to specify the velocity field ${\bf v}$ legitimizes the monitoring of the flow by means of a
series of associated streamlines, just as a direct application of the theory of characteristics \cite{courant-hilbert-bk-2},
although there is no way to connect these paths with the actual paths (if any) quantum particles would follow.
At this level (and to date, with the available experimental data we have), such paths can only be taken as
indicators of how the particles flow on average through configuration space.

Regarding the $Q$ contribution to Eq.~(\ref{eq8}), it is typically called Bohm's quantum potential and has the
functional form of a measure of the curvature of the wave function amplitude,
\be
 Q = - \frac{\hbar^2}{2m} \frac{\nabla^2 A}{A}
    = - \frac{\hbar^2}{4m} \left[ \frac{\nabla^2 \rho}{\rho}
          - \frac{1}{2} \left( \frac{\nabla \rho}{\rho} \right)^2 \right] .
 \label{eq12}
\ee
Its identification with a potential is somehow misleading, since it does not come at all from the potential
energy term in Schr\"odinger's equation, but from the kinetic one.
Notice that, after substituting the ansatz (\ref{eq3}) into the kinetic part of Eq.~(\ref{eq6}) and then
dividing the resulting expression by $\Psi$, we obtain
\be
 - \frac{\hbar^2}{2m} \frac{\nabla^2 \Psi}{\Psi} = \frac{(\nabla S)^2}{2m} + Q
   + \frac{\hbar}{2i} \frac{\nabla \cdot {\bf J}}{\rho} .
 \label{eq13}
\ee
The first two terms provide us with an idea on the system velocity (the first term) and
also with how fast it spreads out (the second term), that is, they constitute some sort of kinetic and
internal energies \cite{sanz:JPA:2008}, respectively, which rule the dynamical behavior of the system state.
The last term (the imaginary one), on the other hand, will contribute to the continuity equation, since it
reflects the conservation of the probability through the divergence of the quantum flux.


\section{Quantum superpositions}
\label{sec3}


\subsection{General aspects}
\label{sec31}

Discussion around Bohmian mechanics can be put at two levels.
One concerns the predictions or calculation of quantum observables.
In this regard, the outcomes rendered by Bohmian mechanics coincide with those provided
by any other standard quantum picture, hence they are also compatible with the experiment.
So far, no difference.
Differences arise in the way how observables are computed.
In standard quantum approaches, the key element is the wave function or, in more general
frameworks, the density operator, which provide us with a global description of the
phenomenon or process described.
That is, we get all accessible information (energy spectra, diffraction patterns, etc.)
in one go.
The Bohmian procedure, on the contrary, is closer in nature to classical-like statistical
approaches (e.g., molecular dynamics or Monte Carlo samplings), where the observable is
reconstructed after collecting a number of data and performing a statistically treatment
over them.
In other words, the observable is obtained event by event.
Since experiments are typically performed in high-intensity regimes, there is
no difference between the standard analysis based on assuming a continuous probability
distribution and the Bohmian one, relying on a granular-like analysis in the large-number
domain --- actually, standard treatments are usually more efficient in this case.
In the low-intensity regime, though, the Bohmian treatment may be quite convenient,
because we can reproduce the experiment on the same event-by-event basis --- even if each
click or spot observed in the experiment cannot be uniquely associated with a single
Bohmian trajectory.

The other level is the dynamical one, i.e., the way how Bohmian mechanics contributes to
understand the evolution or diffusion of quantum probability in the configuration space.
This is done by stressing the role of the quantum
flux and, by means of it, the presence of a dynamical velocity field associated with the quantum
phase, which is not the case in other more standard quantum approaches.
Once we have seen the rudiments of Bohmian mechanics we shall focus on
the equation of motion (\ref{eq10}), which is going to be the key element from now on in this work
concerning the relationship between quantum symmetry and quantum superposition.
In this regard, let us start with the simplest case.
It can easily be seen from the Schr\"odinger equation (\ref{eq6}) that any of its solutions is well determined
except for a constant phase factor.
This means that, if $\Psi$ is a solution, then
\be
 \Psi_\alpha = e^{i\alpha} \Psi
 \label{eq14}
\ee
is also a valid solution for any constant $\alpha$.
Taking into account the definitions (\ref{eq4}) and (\ref{eq5}), we have
\setlength\arraycolsep{2pt}
\ba
 \rho_\alpha & = & \rho ,
 \label{eq15} \\
 S_\alpha & = & S + \alpha .
 \label{eq16}
\ea
Substituting now the relation (\ref{eq16}) into the equation of motion (\ref{eq10}) leads us to the
invariant form
\be
 \dot{\bf r}_\alpha = \dot{\bf r} .
 \label{eq17}
\ee
Accordingly, global constant phase factors are irrelevant regarding the dynamics displayed by the system; the trajectories
associated with $\Psi$ and with $\Psi'$ will be indistinguishable.

\begin{SCfigure}
\centering
  \includegraphics[height=13cm]{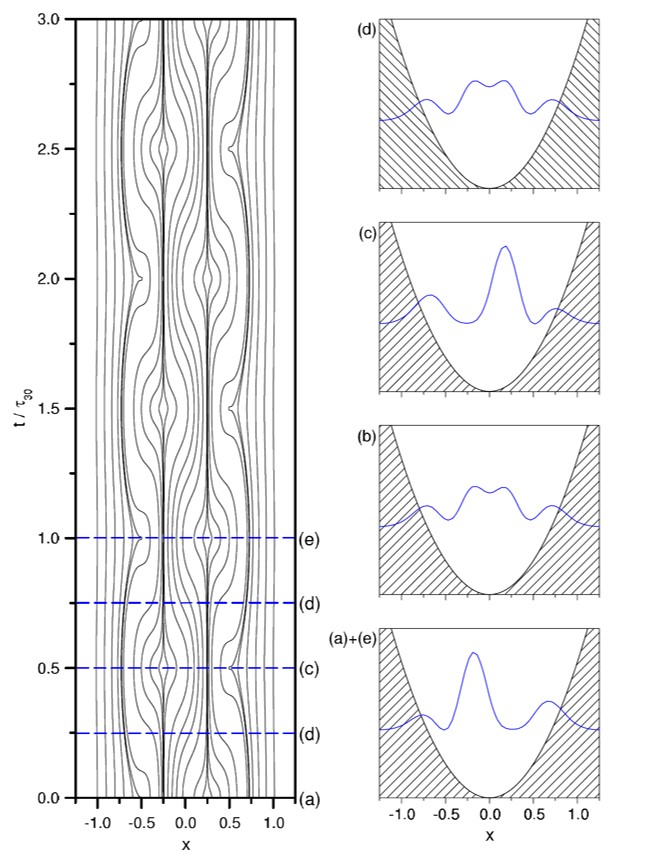}
  \caption{\label{fig2}
   Time-evolution of a two-level state consisting of the superposition of the ground state of the harmonic oscillator
   and its third excited state.
   Sets of Bohmian trajectories (left) are useful to determine the periodic evolution of the flow inside the potential,
   while the usual probability density plots (right) render more general statistical information.
   Comparing both pictures, it can be noticed that, within a cycle [transition from (a) to (e)], swarms of
   (probability) trajectories move towards certain regions while avoiding others, this explaining the appearance of
   maxima in the probability density.
   This reasoning, close to classical statistical ones (in the end, quantum mechanics is a statistical theory), is different
   from the usual one based on the idea of constructive and destructive interference discussed above.
   [For visual purposes, the initial conditions of the trajectories are uniformly distributed along the region covered
   by the initial probability density, $\rho(x,t=0)$, instead of being weighted with this density.])}
\end{SCfigure}

Sometimes the wave function acquires a global time-dependent phase factor, which is necessary in
order to keep the consistence of the solutions to Eq.~(\ref{eq6}) along time.
This is the case, for example, of the time-dependent phase factor developed by Gaussian wave packets over time
(the direct quantum analog of the Gouy phase in Gaussian beam optics \cite{sanz:PRA:2017}), or the geometric or
Pancharatnam-Berry phase developed by energy eigenstates over the course of a cycle (in this case, $\phi$
would correspond to such energy eigenstates, thus being a time-independent solution to Schr\"odinger's
equation).
In general, these wave functions can be recast as
\be
 \Psi ({\bf r},t) = e^{i\varphi (t)} \phi ({\bf r},t) .
 \label{eq18}
\ee
Unlike the relation (\ref{eq14}), here the full term on the right-hand side is a solution ($\phi$ is the part
of the wave function globally affected by the preceding phase factor, but is not a solution itself, in general).
As before, when (\ref{eq18}) is substituted into the definitions (\ref{eq4}) and (\ref{eq5}), we obtain
\setlength\arraycolsep{2pt}
\ba
 \rho_\Psi & = & \rho_\phi ,
 \label{eq19} \\
 S_\Psi & = & S_\phi + \varphi (t) \hbar ,
 \label{eq20}
\ea
where subscripts here denote whether the field is associated with $\Psi$ or with $\phi$.
However, from the relation (\ref{eq20}), again we obtain that the velocity field is invariant in
those cases where a time-dependent phase factor is added to the phase field, since
\be
 \dot{\bf r}_\Psi = \dot{\bf r}_\phi .
 \label{eq21}
\ee
An interesting scenario in this regard is when the wave function displays nodes, around which the phase
undergoes a $2\pi$ jump once a full loop has been closed.
In such cases, although the flux is quantized, the trajectories moving around the nodes are not affected by the
sudden jump in the phase \cite{sanz:jcp:2004,sanz:prb:2004}.
Another example is that of a coherent superposition formed by two energy eigenstates of a bound potential.
Consider, for an instance, the superposition of the ground state and the third excited state of the harmonic
potential, which can be recast as
\be
 \Psi(x,t) = c_0 \phi_0 (x) e^{-iE_0 t/\hbar} + c_3 \phi_3 (x) e^{-iE_3 t/\hbar}
  = e^{-iE_0 t/\hbar} \left[ c_0 \phi_0 (x) + c_3 \phi_3 (x) e^{-i\omega_{30}t} \right] .
 \label{eq22}
\ee
As can be seen in Fig.~\ref{fig2}, where a simulation in terms of trajectories (left) and probability
densities (right), only the relative phase factor depending on the frequency
$\omega_{30} \equiv (E_3 - E_1)/\hbar$ is dynamically relevant; the time-dependent prefactor can be
neglected, although it constitutes part of the solution to the Schr\"odinger equation.

\sidecaptionvpos{figure}{c}

Bearing in mind the behavior of the trajectories displayed in Fig.~\ref{fig2} (left panel), before concluding
this section let us consider a general coherent superposition of two states,
\be
 \Psi({\bf r},t) = \Psi_1({\bf r},t) +  \Psi_2({\bf r},t) .
 \label{eq23}
\ee
Each one of the contributing states represents a different possibility for the system, e.g., passing
through one slit or another in a Young-type two-slit experiment, or choosing one path or another in
Wheeler's delayed choice experiment, or just dealing with a two-level system, as in the case described
above.
Consider that each state is recast in polar form, as it was done with the ansatz (\ref{eq3}).
The probability density and the quantum flux can then be recast as
\setlength\arraycolsep{2pt}
\ba
 \rho & = & \rho_1 + \rho_2
  + 2 \sqrt{\rho_1 \rho_2} \cos \mathcal{S} ,
 \label{eq24} \\
 {\bf J} & = & {\bf J}_1 + {\bf J}_2 + 2 \sqrt{\rho_1 \rho_2} \left( \frac{\nabla \bar{S}}{m}
  - \frac{\mathcal{Q}}{\hbar} \tan \mathcal{S} \right) \cos \mathcal{S} ,
 \label{eq25}
\ea
where the definitions (\ref{eq4}) and (\ref{eq11}) have been taken into account, respectively, as well as
the new quantities
\setlength\arraycolsep{2pt}
\ba
 \bar{S} & = & \frac{S_1 + S_2}{2} ,
 \label{eq26} \\
 \mathcal{S} & = & \frac{S_1 - S_2}{\hbar} ,
 \label{eq27} \\
 \mathcal{Q} & = & \frac{Q_1 - Q_2}{2} .
 \label{eq28}
\ea
From these expressions, the associated velocity field reads as
\be
 \dot{\bf r} = {\bf v} =
  \frac{\rho_1}{\rho}\ \! {\bf v_1} + \frac{\rho_2}{\rho}\ \! {\bf v_2}
  + 2 \sqrt{\frac{\rho_1 \rho_2}{\rho^2}} \left( \frac{\nabla \bar{S}}{m}
  - \frac{\mathcal{Q}}{\hbar} \tan \mathcal{S} \right) \cos \mathcal{S} .
 \label{eq29}
\ee
According to this expression, the superposition generates dynamics where each partial wave
contributes with a given weight, $\rho_i/\rho$, with $i=1,2$.
There is also an extra term containing the averaged phase field and the difference between quantum
potentials that makes the dynamics highly nonlinear.
It is by virtue of this nonlinearity that, for instance, in Young's experiment we can eventually
observe interference fringes, which would never arise if only the former weighted sum would
contribute \cite{sanz:EPJD:2007,sanz:CPL:2009-2,luis:AOP:2015}.
Notice that although Eq.~(\ref{eq29}) has been derived or taken under a Bohmian view, it is just standard
and simple quantum mechanics, without the artificial prejudice that trajectories are incompatible with
quantum mechanics or, if trajectories are allowed, that they are associated with (the also artificial)
idea of particle-wave duality.

Next we are going to discuss the physics involved in some particular illustrative examples where the
above superposition holds.
Further details related either to the associated models or to computational aspects can be found in
the references provided in there).


\subsection{Young's two-slit experiment}
\label{sec32}

Apparently, we all may agree that, at least on a formal level, Young's two-slit experiment is well understood,
even if the experiment itself is still considered the mystery having ``in it the heart of quantum mechanics'',
quoting Feynman \cite{feynman:FLP3:1965}.
The situation changes, though, if it is reexamined from a Bohmian perspective, that is, if we switch from one
quantum picture to another.
Although this may sound a bit odd, the hydrodynamic language of Bohmian mechanics is able to highlight a kind
of symmetry that is not apparent or easy to see with the other pictures (although, as mentioned above, it could be inferred).
This symmetry is associated with the well-known {\it non-crossing rule} typically associated with Bohmian
mechanics \cite{holland-bk,sanz:JPA:2008}.

\begin{figure}[t]
\centering
  \includegraphics[width=12cm]{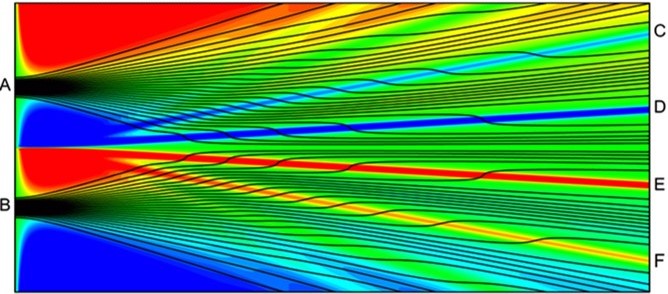}
  \caption{\label{fig3}
   Bohmian explanation of the Young two-slit diagram displayed in Fig.~\ref{fig1}.
   The contour-plot represents the velocity field, with the gradual transition from red to blue denoting the passage
   from lowest values of this field to larger ones.
   As initial condition a sum of two Gaussian wave packets has been considered.
   The trajectories shown (solid lines) illustrate how the quantum probability flows independently emanates from
   each slit and then, after some time, it evolves into a series of separate swarms flowing along the interference
   maxima directions, while avoiding others corresponding to the interference minima (regions C, D, E and F).
   For visual purposes, the initial conditions of the trajectories are uniformly distributed along the regions covered
   by the initial probability density, $\rho(x,t=0)$, instead of being weighted with this density.
   (Further analytical and numerical details can be found in \cite{sanz:EJP-arxiv:2017}.)}
\end{figure}

Consider the Young two-slit experiment illustrated in Fig.~\ref{fig1} in terms of the Huygens' construction.
In brief, in the experimental setup there are two holes or slits on a screen, such that when the matter wave
reaches this screen, it undergoes diffraction at both openings.
The diffracted waves (behind the screen) form a coherent superposition that can be described by
the ansatz (\ref{eq23}).
If the openings are identical, and so the diffracted waves,
Eq.~(\ref{eq29}) forbids trajectories started in the regions covered by each wave packet from crossing the
symmetry line dividing the system into two mirror images \cite{sanz:JPA:2008,sanz:EJP-arxiv:2017}, as can be
seen in Fig.~\ref{fig3}.
The behavior displayed by the trajectories is a manifestation of the above non-crossing rule,
which avoids trajectories from crossing the same point of the position space at the same time.
In this particular simulation the diffracted waves have been described in terms of a coherent superposition
of two identical Gaussian wave packets accounting for the transverse degree of freedom, with each wave packet having
the functional form
\be
 \Psi (x,t) = \left(\frac{1}{2\pi\tilde{\sigma}_t^2}\right)^{1/4}
   e^{-(x - x_t)^2/4\sigma_0\tilde{\sigma}_t + i p (x - x_t)/\hbar + iEt/\hbar} ,
 \label{eq30}
\ee
where $x_t = x_0 + v t$, $p = m v$, $E = p^2/2m$, and
\be
 \tilde{\sigma}_t = \sigma_0 \left( 1 + \frac{i\hbar t}{2m\sigma_0^2} \right) .
 \label{eq31}
\ee
In the case dealt with here, $v = 0$ and the initial position for each wave packet is $x_0 = x_\pm = \pm d/2$,
where $d$ the distance between their centers and $\pm$ makes reference to the relative position of each wave
packet.

The symmetry mentioned above becomes apparent already with tools that we have at hand in quantum mechanics,
as it is the case of the probability density \cite{sanz:EJP-arxiv:2017}.
A close inspection to the evolution of this quantity provide us with information on how the two diffracted beams increase
their size according to the ratio
\be
 \frac{\sigma_t}{\sigma_0} = \sqrt{1 + \left( \frac{\hbar t}{2m\sigma_0^2} \right)^2 }\ ,
 \label{eq32}
\ee
with $\sigma_t = |\tilde{\sigma}_t|$.
At some time, when their width $\sigma_t$ is large enough, they start overlapping and, some time later, their full
overlap gives rise to the appearance of the well-know maxima and minima of interference (modulated by the
corresponding single-slit diffraction pattern \cite{sanz:JPCM:2002}).
Because of the ideal conditions chosen on purpose in this example, the pattern in space and time displayed by the
probability density is going to be symmetric with respect to the center between the two slits ($x = 0$).
Now, the question that we pose here is whether this has any deeper physical implication, and the answer is yes.
It is clear that the probability density by itself is not enough, because it only tells us that there are some directions
along which it is more likely to find or detect the system than others, but still we can appeal to traditional
arguments to explain why, as it was done in Section~\ref{sec1}.
It is precisely a need to go beyond this simple approach why we are going to consider the velocity field defined by
Eq.~(\ref{eq10}) for this particular problem and the trajectories rendered after integration.

In Fig.~\ref{fig3}, the contour-plot represents the evolution of the velocity field along time, with the transition
from red to blue colors denoting the increase from negative to positive velocity values (green denotes nearly zero
velocities).
At $t=0$ the Gaussian wave packet (\ref{eq30}) is real-valued, so a superposition of two of such wave packets
is also real-valued, which implies an initial zero velocity field.
As time proceeds, diffraction causes that both wave packets start developing a phase, and therefore a
relative phase between them starts appearing.
Because both evolve exactly in the same way (notice how to the left in both cases velocities become negative
and to the right, positive), the matching at the center between them becomes a sort of conflict line.
As a consequence, the (quantum) flux associated with one of the wave packets (slits) cannot penetrate the
region dominated by the other, an vice versa, which can be better appreciated when the picture is recast in
terms of a set of streamlines or trajectories.
These trajectories evolve towards the different interference maxima, as expected, but more importantly those
associated with the wave packet labeled with A never cross to the domain of wave packet B, and vice versa.
There is a perfectly symmetric picture for the flux, which has nothing to do with the quantum potential itself
(since it is nearly zero in the far field, as also happens with the velocity field).

Bohmian trajectories themselves are not experimentally accessible, because, as mentioned above,
they are just streamlines along which probability flows rather than corresponding to actual paths followed
by real particles.
If we are talking about particles diffracted by two slits, as in the case illustrated in Fig.~\ref{fig3},
this means that, in principle, we do not know whether
such particles travel along those trajectories, because there is no empirical evidence on that.
However, we know that, on average, a number proportional to $\rho({\bf r},t)$ has passed through a region
$d{\bf r}$ around the point ${\bf r}$ in the position space at a time $t$.
Nonetheless, we {\it do} know that the transverse momentum in that direction is given by $\nabla S({\bf r},t)$,
because this is a property linked to the statistical ensemble dynamics, which is what Schr\"odinger's equation
eventually describes.
Inspired by the Bohmian ideas, an experiment performed at the University of Toronto in 2011 by
Steinberg and coworkers \cite{kocsis:Science:2011} showed that, by using {\it weak measurements} (a slight
perturbation on the quantum system before performing the true measurement, namely a von Neumann
measurement, which helps to obtain complementary information according to Vaidman {\it et al.}\
\cite{aharonov:PRL:1988}), the transverse momentum could be measured on average.
Although the experiment was performed with photons, i.e., zero-mass particles, and Bohmian mechanics, as
it was formerly devised, was employed to describe massive ones (or, in general, any quantum system characterized
by a nonzero mass), the experimental outcomes were in compliance with the expected transverse momentum,
although this momentum is associated with Poynting's vector
(as formerly suggested by Prosser \cite{prosser:ijtp:1976-1} back in 1976)
rather than with Bohm's momentum, and from them an ensemble of paths could be reconstructed, in good agreement
with the corresponding trajectories \cite{sanz:PhysScrPhoton:2009,sanz:AnnPhysPhoton:2010,sanz:JRLR:2010,sanz:EPN:2013}.


\subsection{General counter-propagating wave packets}
\label{sec33}

The system analyzed in the previous section is a singular case of a more general scenario characterized
by superpositions like (\ref{eq23}).
In particular, for the purpose here, let us consider an initial superposition of two Gaussian wave packets like (\ref{eq30}),
\be
 \Psi_0 (x) = c_1 e^{-(x + d/2)^2/4\sigma_{0,2}^2 + i v (x + d/2)/\hbar}
                   + c_2 e^{-(x - d/2)^2/4\sigma_{0,1}^2 - i v (x - d/2)/\hbar} ,
 \label{eq33}
\ee
which represents two counter-propagating wave packets initially moving towards each other.
In the case of Young's two-slit experiment, we set $v=0$ for simplicity.
Thus, only the time-evolution (spreading) of the wave packets was responsible for the generation of phase factors
[see the complex spreading rate (\ref{eq31})] that eventually led to interference traits.
In the case described by the superposition (\ref{eq33}) there are going to be two competing phase factors \cite{sanz:JPA:2008},
one coming from the spreading, and another one from the propagation speeds.
Actually, the latter implies the presence of a (space-dependent) relative phase even at $t=0$,  unlike the two-slit scenario.
Taking this into account, the speeds of the wave packets, $v$, can be selected in such a way that they cross each other and
reach positions opposite to their initial ones with an almost negligible spreading.
For instance, if $x_{f,1} = x_{0,2} = d/2$, the time lasted in going from $x_{0,1}$ to $x_{f,1}$ is $t = d/v$.
Thus, using the expression (\ref{eq32}), if we wish the Gaussian wave packets only increase their widths in about 10\%
when moving from the initial position to the opposite one (i.e., $\sigma_t = 1.1 \sigma_0$), the speed has to be
\be
 v \approx 2.2 \times \left(\frac{\hbar d}{2m\sigma_0^2}\right),
 \label{eq34}
\ee
or, equivalently,
\be
 \frac{v}{v_s} \approx 2.2 \times \left(\frac{d}{\sigma_0}\right) ,
 \label{eq35}
\ee
where $v_s = \hbar/2m\sigma_0$ is the speed or rate at which the wave packet spreads out \cite{sanz:JPA:2008}.
Putting some more numbers, if $\sigma_0 \approx d/20$, in order to ensure a negligible initial overlapping, then $v/v_s \approx 44$,
which gives an idea of the slow wave-packet spreading along the full evolution of the system.
When this conditions are fulfilled, interference is a temporary feature localized around the region where the wave packets
cross each other (no traces of interference remain later on).
Now, is this picture of crossing wave packets correct, even if the description of the phenomenon requires it?

Such has always been the usual approach to superpositions: the superimposed waves are distinguishable,
because each contributing state is itself a separated solution of Schr\"odinger's equation,\footnote{This notion
of distinguishability (or indistinguishability) should not be confused with the one applied to quantum
particles (states), defined in terms of complete sets of communing observables.} as in Eq.~(\ref{eq2}), with a
well-known time-evolution.
However, according to what we have seen in the previous section, that idea may become misleading.
Of course, in the case of Young's experiment it is difficult to determine what is going on after the two wave
packets have reached a rather large spreading and fully overlapped.
In the case dealt with here, though, interference is localized in both space and time, as said above; before and afterwards
we can perfectly distinguish the propagation of both wave packets.
If we appeal to the same picture, as before, we observe again the same kind of symmetry.
Accordingly, the quantum flux associated with one wave packet cannot cross to the region or domain of the
other, and vice versa.
From this, an interesting alternative picture arises:
the outgoing wave packets after interference are not the crossed wave packets,
but the initial ones after having bounced backwards (at the region where interference
took place).
And this is so regardless of the shape of the wave packets, their propagation speeds, or their relative weights in
the superposition.
Nevertheless, notice that this is just a picture about the quantum flux, which, in the end of the day, is not
an observable quantity \cite{heisenberg:ZPhys:1925}
In other words, although the flux may be a convenient tool to ascertain certain particular properties of the theory,
any further physical interpretation must be taken with care, since only a precise experiment can validate it.
For example, in the experiment described at the end of Sec.~\ref{sec32}, the trajectories cannot be associated
with actual photon paths, because the only observable or measured quantities are densities, from which the
transverse flow is (indirectly) obtained.
Anyway, even if it is only at a formal level, if we consider the initial superposition (\ref{eq33}), the following
cases of interest can be specified:
\begin{itemize}
 \item If the two counter-propagating wave packets have the same initial spreading ($\sigma_{0,1} = \sigma_{0,2}$) and weight in the
  superposition ($|c_1|^2 = |c_2|^2 = 1/2$), their speeds are exchanged after interference, in analogy to the transfer of momentum
  between two classical particles of mass $m$ in a context of elastic scattering.

 \item Analogously, if the wave packets have different initial spreadings ($\sigma_{0,1} \neq \sigma_{0,2}$), but still the
  same weight in the superposition, they exchange their corresponding spreadings, giving the impression that they have
  crossed each other, although this is only in appearance.
  What actually happens is that the corresponding fluxes undergo a rearrangement (the trajectories
  associated with the wave packet with smaller spreading have dispersed more, while the trajectories associated
  with the wider wave packet have get closer).
  Again, this a case analogous to momentum transfer in a classical elastic scattering process between two
  particles with different masses.

 \item If the wave packets are not properly normalized or their weights are different (notice that both
  conditions are equivalent), part of the trajectories associated with one of the wave packets (the more
  prominent one in the superposition) pass, after interference, to form part of the ensemble of trajectories
  associated with the other wave packet.
  In this case, there is a gain of trajectories by one of the wave packets and a loss by the other, in analogy
  to a classical case of inelastic scattering.
\end{itemize}

Such behaviors are expected to survive even in those conditions where the action of the quantum potential is
somehow canceled out by removing its contribution within Schr\"odinger's equation.
This procedure leads to a nonlinear Schr\"odinger equation proposed by Nathan Rosen back in the mid-1960s
\cite{rosen:AJP-1:1964,rosen:AJP-2:1964,rosen:AJP:1965},
\be
 i\hbar \frac{\partial \Psi}{\partial t} = - \frac{\hbar^2}{2m}\ \! \nabla^2 \Psi + \left( V - Q \right) \Psi ,
\ee
which gives rise to a Hamilton-Jacobi equation without the contribution of the quantum potential.
This equation has recently been explored in the context of wave-packet superposition \cite{richardson:PRA:2014,chou:AOP:2016},
showing that even in the limit $\hbar \to 0$ classical mechanics is not fully recovered, but the above symmetries persist.
This behavior, already discussed by Rosen in the 1980s \cite{rosen:FoundPhys:1986}, has also been commented by
others, like Berry \cite{berry:LesHouches:1991}, in the context of the standard formulation of quantum mechanics.
To understand this behavior in simple terms, notice that the evolution of wave functions take place in spaces
of reduced dimensionality, that is, either described by positions or described by momenta.
When the above equation is recast in the form of a continuity equation and an apparently classical Hamilton-Jacobi
equation, the same property is maintained.
That is, in the transformation from $\Psi$ to $(\rho,S)$ there is never a condition that implies unfolding positions
and momenta, and therefore breaking the non-crossing rule.
In classical mechanics, on the contrary, positions and momenta are independent variables, which specify an
expanded space, namely the phase space.

Taking into account the above facts, one may wonder whether the analogy with a classical scattering
process can be further exploited.
And the answer is positive.
In the classical scattering process one may simplify the description by appealing to the physics of an
effective mass (the full-system reduced mass) acted by an also effective force (e.g., a central force when
dealing with masses or charges).
In the problem of interference (the fully symmetric case, to simplify) an analogous approach can be considered
if we notice that the collision of a single wave packet with a hard (infinite) wall renders, at the time of
maximal proximity to the potential, an interference pattern analogous to that displayed by the two
counter-propagating wave packets at the time of maximal interference.
The only difference between both cases is that in the single wave-packet problem all maxima have the same
width, while in the two wave-packet problem, if we divide it by a half (along the symmetry line), one of the
maxima has have width.
To circumvent this problem, we may assume that the collision is not just against a hard wall, but that there
is a well close to the wall, such that part of the wave packet may get attached to it after the collision,
with the corresponding resonance have the appropriate width, i.e., half the width of the interference
maxima.

\begin{figure}[t]
\centering
  \includegraphics[width=13cm]{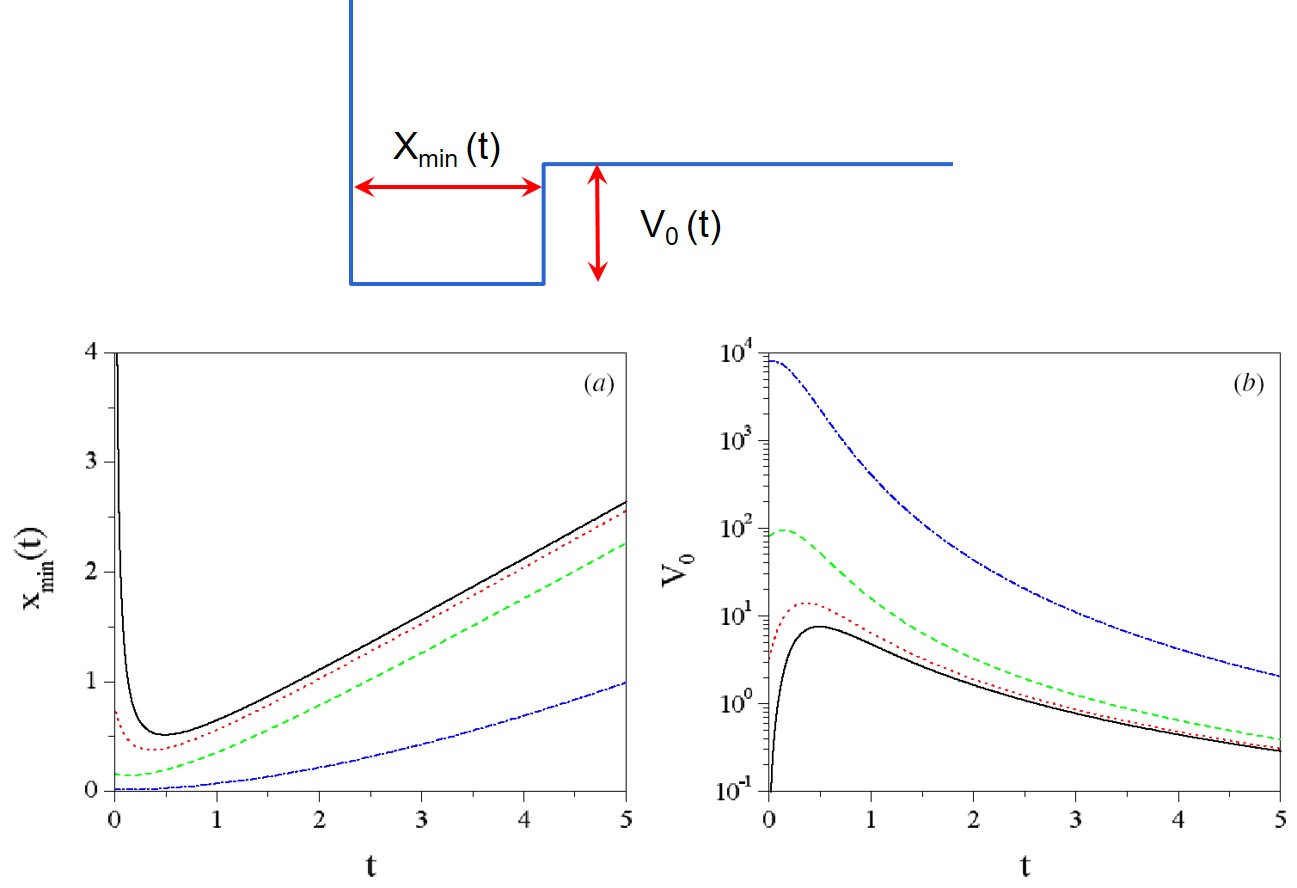}
  \caption{\label{fig4}
   Simple toy-model that allows an effective substitution of a coherent superposition of two wave packets by the scattering
   of a single wave packet with a hard wall followed by a dynamical attractive well with time-dependent
   width and depth (both as being functions of the parameters defining the wave packets) \cite{sanz:JPA:2008}.
   Such time-dependence is explicitly shown in the two panels below for four different cases with increasing value
   of the propagation speed of the wave packets.
   The black solid like denotes a scenario recreating Young's two-slit experiment, while the blue dashed-dotted
   line indicates just the opposite regime, namely a scenario where the wave-packet spreading is negligible compared
   to the propagation speed.
   (Further analytical and numerical details can be found in \cite{sanz:JPA:2008}.)}
\end{figure}

A simple toy model that gathers such features can be constructed \cite{sanz:JPA:2008}, where
the interference between two different wave packets is {\it effectively} described by the (self)interaction
of a single wave packet with its own reflection.
This model, sketched in Fig.~\ref{fig4}, consists of a hard repulsive wall followed by square-well attractive region,
with both the width and depth of the well being time-dependent:
\be
 V(t) = \left\{
  \begin{array}{ccl}
   0      & \qquad & x < x_{\rm min}(t) \\
  -V_0[x_{\rm min}(t)]  & \qquad & x_{\rm min}(t) \le x \le 0 \\
  \infty  & \qquad & 0 < x
  \end{array} \right. ,
 \label{eq36}
\ee
where
\setlength\arraycolsep{2pt}
\ba
  V_0[x_{\rm min}(t)] & = & \frac{2\hbar^2}{m} \frac{1}{x_{\rm min}^2(t)} ,
 \label{eq37} \\
  x_{\rm min}(t) & = & \frac{\pi}{\displaystyle \frac{2p}{\hbar}
  + \frac{\hbar t}{2m\sigma_0^2}\frac{x_t}{\sigma_t^2}}
  = \frac{\pi\sigma_t^2}{\displaystyle \frac{2p\sigma_0^2}{\hbar}
   + \frac{\hbar t}{2m\sigma_0^2} \ \! x_0} ,
 \label{eq38}
\ea
with $p = mv$.
The time-dependence in the width and the depth of the well is precisely related to the fact that the resonant
state has to asymptotically increase its width and, at the same time, the shallowness will decrease.
These two quantities are represented in the two lower panels of Fig.~\ref{fig4} for different values of the
propagation speed of the wave packets, where the black solid line denotes a typical Young-like regime and
the blue dashed-dotted line a case with negligible dispersion.
As can be noticed, in the Young-like regime, the minimum width of the well is rather large, which means that
the wave packet is fully inside the attractive well, while in the case with negligible dispersion the wave
packet starts outside it.
This already gives a clue on why in one case interference will be a persisting trait, while in the other it
is only temporary (and spatially localized).
Nevertheless, also notice that for the Young-type scenario the potential is rather shallow, while in the other
scenario the depth is remarkable.
This sort of play between width and depth can then be rather misleading: apparently meaningless potential wells
may lead to important quantum effects (persistent interference), while the opposite is not necessarily true (remember
that the scenario with negligible dispersion can be well approximated by a typical ray-based description, as it
is commonly done when explaining atomic Mach-Zehnder interferometers \cite{keith:PRL:1991,cronin:RMP:2009} or
Wheeler's delayed choice experiment \cite{wheeler:1978}).

Although such phenomena have not been experimentally confirmed yet, as it is the case of the fluxes in Young's
experiment, it is worth noting that they constitute a key step in matter-wave interferometry, where an incident
matter wave is split up and then, the two partial waves, are made to overlap again on some spot, as it happens
in atomic Mach-Zehnder interferometers \cite{sanz:AOP:2015} or in the renowned Wheeler delayed-choice
experiment \cite{sanz:foundphys:2015} when both setups are left in open configuration.
Although we are not going to enter into details (the interested reader is kindly addressed to the
corresponding references), it is worth noting that in either case the spreading of the two split wavefronts
is negligible with their passage through the interferometer and that the performance is due to a broken
or restored phase symmetry at the end of the interferometer, by disregarding or inserting an element (a
grating in the first case and and a beam splitter in the second case) that is able to undo the phase
splitting at the entrance of the interferometer by precisely an analogous element.


\subsection{Grating diffraction and quantum carpets}
\label{sec34}

\begin{SCfigure}
\centering
  \includegraphics[height=16cm]{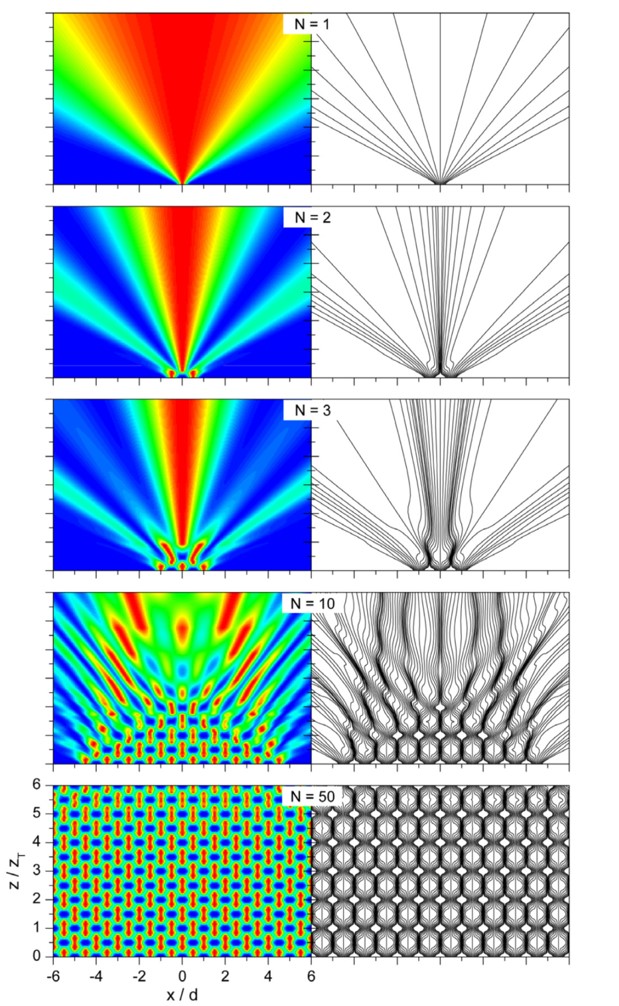}
  \caption{\label{fig5}
   Emergence of a Talbot carpet as the number of slits (diffracted beams) increases.
   The probability density along the transverse direction ($x$) as a function of the longitudinal coordinate [$z$, normalized to
   twice the Talbot distance, Eq.~({eq39})] is displayed in the form of contour-plots on the left column.
   The gradual transition from blue to red denotes increasing values of the density; for a better visualization, at any $z$ value the
   maximum of the probability density has been normalized to unity.
   The Bohmian trajectory picture for each case is displayed on the corresponding right-column panels.
   Also for visual purposes, the initial conditions of the trajectories are uniformly distributed along the regions covered
   by the initial probability density, $\rho(x,z=0)$, instead of being weighted with this density.
   (Further analytical and numerical details can be found in \cite{sanz:JCP-Talbot:2007}.)}
\end{SCfigure}

So far we have discussed the behavior of quantum systems described by two-state superpositions in the position
representation (remember the two-state superposition with energy eigenvalues of the harmonic oscillator from
Section~\ref{sec31}), with Young's two-slit experiment being a paradigm of such systems and counter-propagating
wave packets a generalization of such a case.
What happens if we increase the number of slits or, equivalently, the number of contributing waves to the coherent
superposition, trying a general superposition like (\ref{eq2})?
This is precisely the case shown in Fig.~\ref{fig5}, where we go from a single wave packet to a large
number of them in order to illustrate a transition from single-slit diffraction to diffraction by a $N$-slit grating
\cite{sanz:JCP-Talbot:2007} (as before, we only focus on the transverse direction).
As can be noticed, with increasing number of wave packets the emergence of a highly ordered pattern or structure
becomes more and more apparent in the near field or Fresnel regime, i.e., the space region behind an aperture or
obstacle where the matter wave still strongly feels the action of such a diffracting element, and that is characterized
by a rearrangement or quantization of the momenta that later on, far away from the diffracting element, will give
rise to the Fraunhofer features in the position representation \cite{sanz:JPCM:2002,sanz:AJP:2012}.
Such an emergent structure is called a {\it Talbot carpet} in analogy to the so-called Talbot phenomenon in optics
\cite{talbot:PhilosMag:1836}, and that has direct atomic analogs \cite{pritchard:PRA:1995}.

\sidecaptionvpos{figure}{c}

In order to analyze the Talbot structure with matter waves, here we are also going to proceed as in previous sections
and consider only the transverse coordinate in the description for practical purposes.
So, if the distance between centers of slits (which are assumed to be characterized by a Gaussian transmission
function \cite{sanz:AOP:2015}) is $d$, the grating is perfectly periodic (infinite extension along $x$), and the (de
Broglie) wavelength of the incident matter wave is $\lambda$, an exact replica of the initial diffracted wave can be
found at integer multiples of the distance
\be
 2z_T = \frac{2d^2}{\lambda} ,
 \label{eq39}
\ee
where $z_T$ is called the {\it Talbot distance}.
This distance is, actually, the one at which we can observe a full replica of the initial wave, although it is half a period
of the grating (i.e., $d/2$) shifted with respect to the latter, which means that there are three space and time translational
symmetries:
\setlength\arraycolsep{2pt}
\ba
 \Psi(x + \ell d, t) & = & \Psi (x,t) ,
 \label{eq40} \\
 \Psi(x, t + \tau_T) & = & \Psi(x,t) ,
 \label{eq41} \\
 \Psi(x + d/2, t + \tau_T/2) & = & \Psi(x,t) ,
 \label{eq42}
\ea
with $\ell = \pm 1, \pm 2, \ldots$, and where the recurrence time $\tau_T = md^2/\pi\hbar$ is the {\it Talbot time}.
Apart from this evident symmetry, by increasing the period of the grating it would also possible to observe other
fractional recurrences \cite{arndt:OptExpress:2009,cronin:NJP:2009}, although we are not going to enter into details
in this regard, since they will be discussed later on when dealing with the problem of bound diffraction in next section.

As in previous sections, we are going to switch to the trajectory picture to elucidate whether the symmetries
displayed by the Talbot carpet have further physical implications (although at this point it is already clear they
will have).
To this end, it is useful to define the domain of a slit in analogy to the concept of unit cell in solid state physics
or condensed matter \cite{ashcroft-bk}, which in the present context is understood as the region spanning a distance $d$ around the center
of a given slit [if the center of the slit is $x_0$, the slit unit cell spans the region $x \in [x_0 - d/2, x_0 + d/2)$].
Proceeding this way, we readily notice that the translational symmetry of the carpet gives rise to a kind of unbound
channeling structure: although there are no physical boundaries at all (coming from a potential function or even the
quantum potential, the trajectories started in the domain of a given unit cell keep evolving in a way analogous to being
confined within an infinite box, without crossing to adjacent unit cells.
In other words, the whole pattern can be reconstructed by repeating exactly the same structure, which provides
a physical picture to a well-known result in solid state theory, namely Bloch's theorem, as well as a physical
interpretation in the position representation for the Born-von Karman boundary conditions \cite{ashcroft-bk}, thus
setting a bridge between two different disciplines, as they are the solid state physics (typically understood in the
reciprocal momentum and energy representations) and the matter-wave interferometry or matter-wave optics (typically
explained within the position representation).
As can be noticed, in the problem of Talbot carpets the wave function behind the grating can be described as a
result of the linear combination of a series of diffracted waves.
However, the fact that there is a periodicity implies that, if such diffracted waves are decomposed in terms of
plane waves (with well-defined energies and momenta), most components must be neglected, only remaining
a countably infinite set of allowed plane waves.
In other words, periodicity in this context is equivalent to quantize the momentum (and energy), as also happens in
solid state physics with Bloch's waves and the appearance of energy bands after introducing periodic boundary conditions.
This can be better appreciated by means of a simple example.
Consider that the state describing the diffraction through the slits of the grating is given by Gaussian wave
functions like (\ref{eq30}).
Applying the translational invariance condition (\ref{eq40}), such wave function becomes \cite{sanz:JCP-Talbot:2007}
\be
 \Psi(x,t) = \sqrt{\frac{1}{d}} \left( \frac{8\pi\sigma_0^2}{d^2} \right)^{1/4}
  \sum_{|n|=0}^\infty e^{-\sigma^2 p_n^2/\hbar^2 + ip_n x /\hbar - i\omega_n t} ,
 \label{eq43}
\ee
where $p_n = 2\pi n\hbar/d$ and $\omega_n = 2\pi^2 n^2\hbar^2/md^2$ are quantized quantities.
Notice that, although the analysis has been restricted to what happens within a single period of the grating, the fact
that the wave function (\ref{eq43}) is given in terms of planes waves serves to replicate the behavior within a single
unit cell all the way through the whole grating.
If the number of slits does not go to infinity, though, then the pattern is restricted to a certain space area, showing
an approximately triangular shape, as can be seen in Fig.~\ref{fig5} for low $N$ (for large $N$ we need to go beyond
the frames shown in the figure); beyond this area the typical Fraunhofer grating
diffraction traits start appearing, which is precisely what we observe in the far field in atomic Mach-Zehnder interferometers \cite{sanz:AOP:2015},
but also in diffraction by reflection gratings, as it is the case of atom scattering off periodic surfaces
\cite{sanz:prb:2000,sanz:EPL:2001}.
In such cases, Eq.~(\ref{eq43}) cannot be used and the exact form for the full superposition has to be considered.

\begin{figure}
\centering
  \includegraphics[width=14cm]{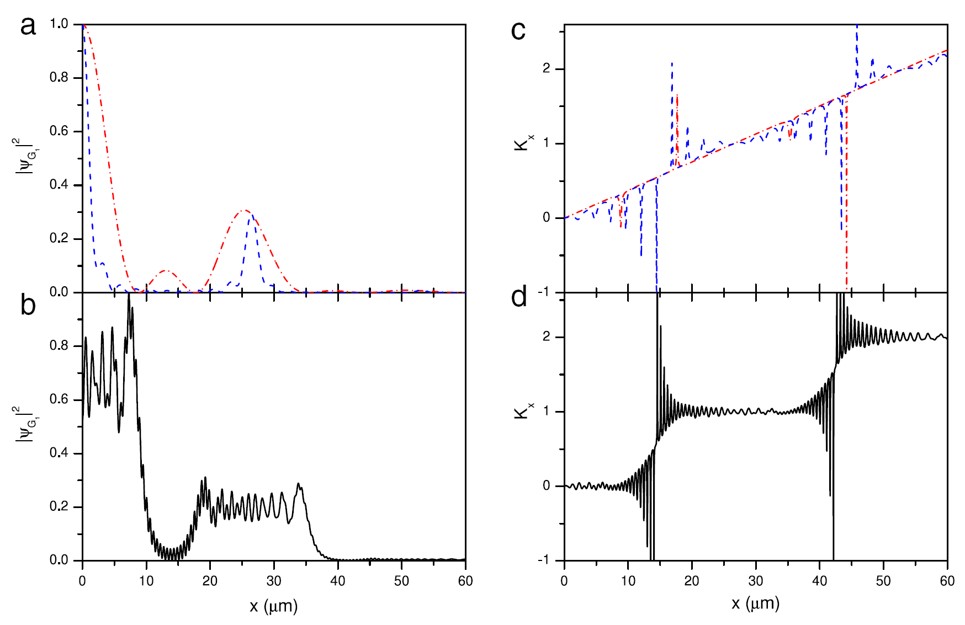}
  \caption{\label{fig6}
    As the number of slits considered to numerically recreate the diffracted wave in an atomic Mach-Zehnder interferometer, just before
    getting diffracted again by a second grating, it observed how a few, distinctive principal interference maxima in the the probability density
    along the transverse direction become something closer to square waves, each progressing along a particular diffraction direction (in
    scattering off surfaces or lattices these directions correspond to the Bragg angles \cite{sanz:prb:2000}).
    In panel (a), the blued dashed line and the red dashed-dotted one represent the probability density for 3 and 11 slits, respectively, good
    examples of a few-slit regime.
    On the contrary, the many-slit regime is represented in panel (b) with 51 slits.
    In the right-column panels, the corresponding transverse (Bohmian) momentum has been represented for exactly the same cases.
    As it can be noticed, in the few-slit regime a series of spikes appear coinciding with the minima of the density, analogous to what can be
    observed in Young's experiment \cite{kocsis:Science:2011,luis:AOP:2015}.
    This structure evolves into an approximate step-ladder one as the number of slits becomes larger and larger, as seen in panel (d), with
    each step indicating a quantized value of the transverse momentum (a Bragg direction).
    For a better understanding of this phenomenon, the transverse momentum is normalized to $2\pi/d$.
   (Further analytical and numerical details can be found in \cite{sanz:AOP:2015}.)}
\end{figure}

Now, apart from such important physical consequences concerning the application of these ideas to the description and
understanding of matter-wave diffraction processes and the analysis of different types of matter-wave interferometry
techniques and their outcomes \cite{sanz:AOP:2015}, the approach can be smartly used to devise and implement
efficient numerical tools on an event-by-event basis.
Observe that the calculation of the trajectories does not require either considering the full expression for the
superposition with all Gaussians, but just (\ref{eq43}), since the periodicity along the full transverse direction is
already contained within this relation in a compact form (all the dynamical information encoded in a single
Gaussian).
This information is automatically transferred to the Bohmian trajectories.
Thus, substitution of the ansatz (\ref{eq43}) into the equation of motion (\ref{eq10}) renders
\be
 \dot{x} =  \frac{\sum_{i,j} p_i e^{-\sigma_0^2 (p_i^2 + p_j^2)/\hbar^2}
    \cos [ (p_i - p_j)x/\hbar - (\omega_i - \omega_j)t ]}
   {\sum_{i,j} e^{-\sigma_0^2 (p_i^2 + p_j^2)/\hbar^2}
    \cos [ (p_i - p_j)x/\hbar - (\omega_i - \omega_j)t ]}
 \label{eq44}
\ee
with $i, j = \pm 1, \pm 2, \ldots$
This equation admits a similar analysis to Eq.~(\ref{eq29}), from which one gets an interesting answer: the periodic boundary conditions
translate, in the trajectory picture, in a sort of impenetrable walls that constrain the evolution of the trajectories to a series of channels.
Only if the number of openings is finite, such channels will get diluted far from the grating and will eventually disappear, letting the trajectories
to start gathering towards the different Fraunhofer interference features.
It can be shown \cite{sanz:JCP-Talbot:2007,sanz:AOP:2015} that in the far-field limit, for a relatively large number of openings, the flux is
quantized and acquires a ladder shape, with each step being associated with a diffraction order, as seen in Fig.~\ref{fig6} --- if the number of
openings considered is rather small (see upper panels of Fig.~\ref{fig6}), then the structure of the transverse quantum flux is closer to what
Steinberg and coworkers observed in the case of photons \cite{kocsis:Science:2011,luis:AOP:2015}.

\subsection{Bound diffraction}
\label{sec35}

In Section~\ref{sec35} we have said (and illustrated with simulations) that the channel structure found when considering perfectly periodic gratings is analogous
to what we find when a true channel is analyzed, i.e., the confinement of a matter wave inside an infinite square potential.
This situation essentially corresponds to diffraction under confinement, where such an analogy can be easily shown \cite{sanz:JCP-Talbot:2007} if instead of
considering the symmetry imposed by the boundary condition (\ref{eq40}), we impose on the Gaussian wave packet the condition that the constituting plane
waves have to be eigenstates of the infinite potential (i.e., solutions that vanish, as well as their first derivatives, at the boundaries of the potential).
In this case, and assuming that the Gaussian is nearly zero close to the borders, we obtain
\be
 \Psi(x,t) = \sqrt{\frac{8}{d}} \left( \frac{2\pi\sigma_0^2}{d^2} \right)^{1/4}
  \sum_{n=0}^\infty e^{-\sigma_0^2 p_n^2/\hbar^2 - i E_n t} \cos(p_n x/\hbar) ,
 \label{eq45}
\ee
where $p_n = (2n+1)\pi\hbar/d$ and $E_n = p_n^2/2m$.
These solutions, analogous to the example displayed in Fig.~\ref{fig2} (although in that case it was for a harmonic oscillator), have full recurrences
whenever the time is an integer multiple of the period
\be
 \tau_r = \frac{2\pi}{\omega_{1,0}} = \frac{2\pi\hbar}{E_1 - E_0} = \frac{md^2}{2\pi\hbar} ,
 \label{eq46}
\ee
which coincides with the time at which we observe the first Talbot recurrence in the grating (the half-a-period shifted), at $\tau_T/2$.
It is worth noting here that the symmetry displayed by the wave function in this case is of the kind described by the relations (\ref{eq19})
and (\ref{eq20}), since
\be
 \Psi (x,t + \tau_r) = e^{i\varphi (t)} \Psi(x,t) .
 \label{eq47}
\ee

\begin{figure}
\centering
  \includegraphics[width=16cm]{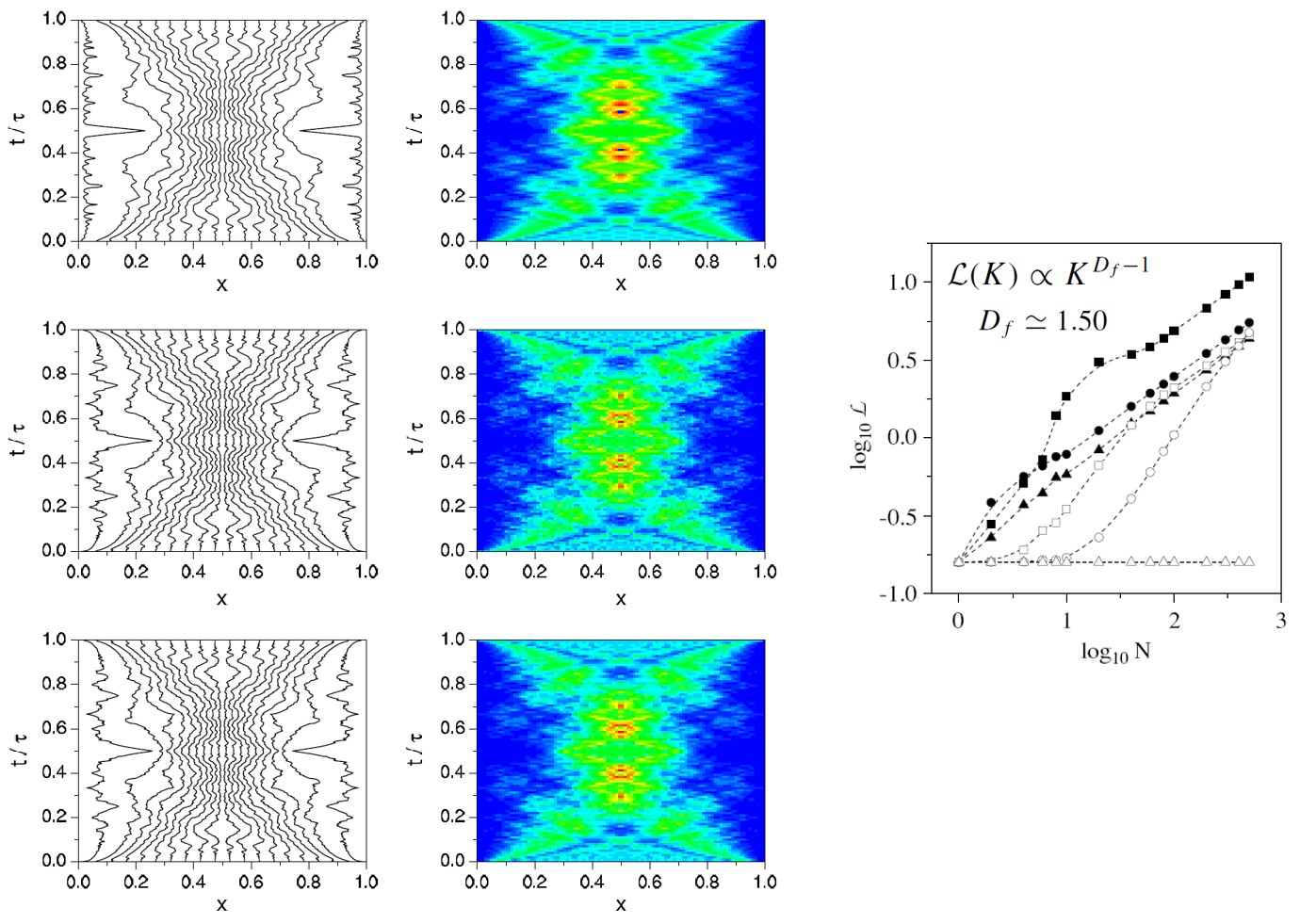}
  \caption{\label{fig7}
   Time-evolution of a square wave function of width $w$ inside an infinite well of length $L > w$.
   The panels on the left and central columns show the evolution of the trajectories and probability density as a function
   of the number of eigenstates considered, increasing from top to bottom.
   As it is apparent, such an increase implies a higher degree of fractality in both the trajectories and the shape displayed
   by the probability density.
   For visual purposes, the initial conditions of the trajectories are uniformly distributed along the regions covered
   by the initial probability density, $\rho(x,t=0)$, instead of being weighted with this density.
   On the right, the panel shows the measure of the length of some selected trajectories as a function of the number of eigenstates involved in
   the initial superposition (both in log scale).
   As the initial conditions of the trajectories become closer to the edge of the initial state, the fractal dimension gets closer to
   that of the probability density --- actually, notice that in the limit of large $N$, all trajectories approach the same fractal dimension.
   (Further analytical and numerical details on the model can be found in \cite{sanz:JPA:2005}.)}
\end{figure}

The type of symmetries or quantum carpets observed in confined systems has a very interesting limit.
If the shape of the initial diffracted wave is a square instead of a Gaussian, as shown by Michael Berry \cite{berry:JPA:1996}, at some times the quantum
state displays fractal features even if these carpets have the same global properties associated with diffracted states inside confined potentials
\cite{sanz:AOP-arxiv:2017}.
These fractal states arise whenever the diffracted wave has sharp edges, i.e., it is continues everywhere, but not differenciable at some points, as it happens
with a square function of width $w$ inside a well of length $L > w$.
As in Section~\ref{sec31}, the wave function can be defined in terms of a linear combination of energy eigenstates,
\be
 \Psi (x,t;N) = \sum_{n=1}^N c_n \phi_n (x) e^{-iE_n t/\hbar} ,
 \label{eq48}
\ee
although here the expectation value of the energy is going to monotonically increase with the number of eigenfunctions considered in the superposition.
In this expression, $\phi_n (x)$ denote the corresponding eigenfunctions and $|c_n|^2$ the relative weights of each component in the superposition (or,
equivalently, the population of each energy level).
It can be shown \cite{sanz:JPA:2005} that the fractal dimension of the quantum state can be obtained from the measure
\be
 \mathcal{L}(\mathcal{K}) \propto \mathcal{K}^{D_f - 1} ,
 \label{eq49}
\ee
where $\mathcal{L}$ denotes the the Euclidean length of, say, the probability density, and $\mathcal{K}$ the number of eigenfunctions considered in the
superposition.
As $\mathcal{K}$ increases, the length $\mathcal{L}$ also increases, and from their relation, in the limit of large $\mathcal{K}$, we obtain the fractal
dimension of the state, $D_f$, which in the particular case displayed in Fig.~\ref{fig7} is approximately 1.50.
The Bohmian trajectories associated with the state (\ref{eq48}) can be obtained by means of the expression
\be
 \dot{x}_N = \frac{\hbar}{m}\ \! {\rm Im}
   \left\{ \frac{1}{\Psi(x,t;N)} \frac{\partial \Psi(x,t;N)}{\partial x} \right\}
 \label{eq50}
\ee
with the limiting case
\be
 x(t) = \lim_{N \to \infty} x_N (t)
 \label{eq51}
\ee
being, in general, a fractal trajectory (if the initial condition does not correspond to the central line of the state) with the same dimension of the quantum
state [in this particular example, $D_f \approx 1.50$, when measured according to (\ref{eq49})].


\section{Final remarks}
\label{sec4}

The starting point of this work was a very specific question: has physical superposition the same
meaning as mathematical superposition?
This question was then reoriented to the presence of symmetries displayed by matter waves under
different circumstances in diffraction-like experiments.
To tackle the issue, apart from using standard quantum techniques and others coming from Fourier
optics (spectral-like decompositions of the wave function), the Bohmian picture of quantum mechanics
has also been taken as a very convenient tool to explore the physics of such phenomena beyond
ordinary, well-known quantum results.
In this picture, the emphasis is put on the quantum flux and the trajectories or streamlines ``flying''
with it, without no detriment to other more usual probability density based considerations.
Notice, though, that the use of such trajectories has been taken as a probing tool to understand the
``hidden'' implications of interference-mediated symmetries, and not with an ontological purpose
(associating them with the actual motion of real quantum particles, i.e., as ``hidden variables'').

Such an analysis renders a series of interesting conclusions, which can be summarized as follows:
\begin{itemize}
 \item The symmetries displayed by the probability density, apart from being mathematically describable
  in terms of superpositions, are associated with distinctive non-superimposing flux dynamics that split up the
  configuration (position) space into separate dynamical domains.

 \item Those domains can be analyzed by means of effective potential models that constrain the evolution
  of the quantum system to them, neglecting what happens everywhere else.
  From these models it is then possible to get a precise and intuitive idea of what spatial coherence means,
  setting the grounds for future experiments to explore the ``weirdness'' of quantum systems and their capability
  to display interference.

 \item From the exchange symmetries displayed by coherent superpositions consisting of two counter-propagating
  wave packets, we have seen that they behave analogously to two classical colliding particles, identifying regimes
  similar to classical elastic and inelastic scattering events.

 \item Superpositions of periodically repeated wave functions generate densities with translational invariance,
  namely Talbot carpets, which are analogous to the structures we find in solid state physics and condensed matter physics
  due to the momentum (and energy) quantization induced on the initial wave function.
  Furthermore, such structures, when analyzed in terms of trajectories, seem to give rise or to be originated by repetitions of
  the dynamical behavior displayed by subsets of trajectories contained within individual domains or unit cells.

 \item The confinement or quantization observed in periodic grating diffraction processes is isomorphic with the behavior
  of matter waves inside bound potentials, including similar recurrence periods and trajectory behaviors.
\end{itemize}

These results, rather than constituting a conclusion from a series of works on matter-wave diffraction, they open new ways
to rethink these processes and design new experiment to test the peculiarities of quantum mechanics.
The experiment carried out by Steinberg and coworkers opened a new path to understand Young's two-slit experiment on
a different level, but also to keep digging down in our understanding of quantum phenomena beyond elementary (and
unrealistic) path-interfering arguments, as commonly done in matter-wave interferometry (Mach-Zehnder or Talbot-Lau) or
to introduce ``paradoxes'', such as Wheeler's delayed-choice experiment.
The quantum reality is far more complex (and richer) than such reasonings, which have to be overcome once for all, particularly taking into
account the technological advances we have currently at hand to investigate it in the laboratory.
Hence it is a matter of going beyond old-fashioned prejudices and using different alternative tools and routes to explore them.
The Bohmian picture is probably a suitable tool to that end, but not necessarily the only one or the best one.
At least, here it has served us to reach a deeper understanding of interference-mediated symmetries in matter-wave
diffraction phenomena.


\ack

The author thanks the organizers of the International Symposium {\it Symmetries in Science XVII} (Bregenz,
Austria, 2017) for their kind invitation to participate in the event.
On a more personal side, he would also like to express his gratitude to Dieter and Ivette for their exquisite
hospitality and attention during this event.
The author also acknowledges financial support from the Spanish MINECO (grant reference number FIS2016-76110-P).


\section*{References}


\begin{thebibliography}{10}
\expandafter\ifx\csname url\endcsname\relax
  \def\url#1{{\tt #1}}\fi
\expandafter\ifx\csname urlprefix\endcsname\relax\def\urlprefix{URL }\fi
\providecommand{\eprint}[2][]{\url{#2}}

\bibitem{young:PTRSL:1804}
Young T 1804 {\em Phil. Trans. R. Soc. Lond.\/} {\bf 94} 1--16

\bibitem{einstein:AnnPhys:1905-1}
Einstein A 1905 {\em Ann. Phys.\/} {\bf 17} 132--148

\bibitem{Shizgal-bk:2015}
Shizgal B 2015 {\em Spectral Methods in Chemistry and Physics: Applications to
  Kinetic Theory and Quantum Mechanics\/} Scientific Computation (Heidelberg:
  Springer)

\bibitem{arndt:NNanotech:2012}
Juffmann T, Milic A, M\"ullneritsch M, Asenbaum P, Tsukernik A, T\"uxen J,
  Mayor M, Cheshnovsky O and Arndt M 2012 {\em Nat. Nanotech.\/} {\bf 7}
  297--300

\bibitem{arndt:NaturePhys:2014}
Arndt M and Hornberger K 2014 {\em Nature Phys.\/} {\bf 10} 271--277

\bibitem{taylor:ProcCambPhilSoc:1909}
Taylor G~I 1909 {\em Proc. Camb. Philos. Soc.\/} {\bf 15} 114--115

\bibitem{jonsson:ZPhys:1961}
J\"onsson C 1961 {\em Z. Phys.\/} {\bf 161} 454--474

\bibitem{jonsson:AJP:1974}
J\"onsson C, Brandt D and Hirschi S 1974 {\em Am. J. Phys.\/} {\bf 42} 4--11

\bibitem{pozzi:AJP:1973}
Donati O, Missiroli G~F and Pozzi G 1973 {\em Am. J. Phys.\/} {\bf 41} 639--644

\bibitem{pozzi:AJP:1976}
Merli P~G, Missiroli G~F and Pozzi G 1976 {\em Am. J. Phys.\/} {\bf 44}
  306--307

\bibitem{tonomura:ajp:1989}
Tonomura A, Endo J, Matsuda T, Kawasaki T and Ezawa H 1989 {\em Am. J. Phys.\/}
  {\bf 57} 117--120

\bibitem{sanz:JPA:2008}
Sanz A~S and Miret-Art\'es S 2008 {\em J. Phys. A: Math. Theor.\/} {\bf 41}
  435303(1--23)

\bibitem{sanz:AOP:2015}
Sanz A~S, Davidovi\'c M and Bo\v{z}i\'c M 2015 {\em Ann. Phys.\/} {\bf 353}
  205--221

\bibitem{sanz:SSR:2004}
Guantes R, Sanz A~S, Margalef-Roig J and Miret-Art\'es S 2004 {\em Surf. Sci.
  Rep.\/} {\bf 53} 199--330

\bibitem{ashcroft-bk}
Ashcroft N~W and Mermin N~D 1976 {\em Solid State Physics\/} (Fort Worth, USA:
  Harcourt College Publishers)

\bibitem{bohm:PR:1952-1}
Bohm D 1952 {\em Phys. Rev.\/} {\bf 85} 166--179

\bibitem{bohm:PR:1952-2}
Bohm D 1952 {\em Phys. Rev.\/} {\bf 85} 180--193

\bibitem{holland-bk}
Holland P~R 1993 {\em The Quantum Theory of Motion\/} (Cambridge: Cambridge
  University Press)

\bibitem{sanz:EJP-arxiv:2017}
Sanz A~S 2018 {\em Front. Phys.\/} (in press, 2018);
\textit{preprint} \eprint{arXiv:1707.00609v1\/} {\bf [quant-ph]}
1--13

\bibitem{sanz:JPhysConfSer:2014}
Sanz A~S 2014 {\em J. Phys.: Conf. Ser.\/} {\bf 504} 012028(1--14)

\bibitem{pozzi:EJP:2013}
Matteucci G, Pezzi M, Pozzi G, Alberghi G~L, Giorgi F, Gabrielli A, Cesari N~S,
  Villa M, Zoccoli A, Frabboni S and Gazzadi G~C 2013 {\em Eur. J. Phys.\/}
  {\bf 34} 511--517

\bibitem{madelung:ZPhys:1926}
Madelung E 1926 {\em Z. Phys.\/} {\bf 40} 322--326

\bibitem{schiff-bk}
Schiff L~I 1968 {\em Quantum Mechanics\/} 3rd ed (Singapore: McGraw-Hill)

\bibitem{morse-bk-2}
Morse P~M and Feshbach H 1953 {\em Methods of Theoretical Physics\/} vol~2 (New
  York: McGraw-Hill)

\bibitem{furth:ZPhys:1933}
F\"urth R 1933 {\em Z. Phys.\/} {\bf 81} 143--162

\bibitem{comisar:PhysRev:1965}
Comisar G~G 1965 {\em Phys. Rev.\/} {\bf 138} B1332--B1337

\bibitem{courant-hilbert-bk-2}
Courant R and Hilbert D 1966 {\em Methods of Mathematical Physics\/} vol~2 (New
  York: John Wiley \& Sons)

\bibitem{sanz:PRA:2017}
Sanz A~S, Davidovi\'c M and Bo\v{z}i\'c M submitted, 2017 {\em Phys. Rev. A\/}
  1--15

\bibitem{sanz:jcp:2004}
Sanz A~S, Borondo F and Miret-Art\'es S 2004 {\em J. Chem. Phys.\/} {\bf 120}
  8794--8806

\bibitem{sanz:prb:2004}
Sanz A~S, Borondo F and Miret-Art\'es S 2004 {\em Phys. Rev. B\/} {\bf 69}
  115413(1--5)

\bibitem{sanz:EPJD:2007}
Sanz A~S and Borondo F 2007 {\em Eur. Phys. J. D\/} {\bf 44} 319--326

\bibitem{sanz:CPL:2009-2}
Sanz A~S and Borondo F 2009 {\em Chem. Phys. Lett.\/} {\bf 478} 301--306

\bibitem{luis:AOP:2015}
Luis A and Sanz A~S 2015 {\em Ann. Phys.\/} {\bf 357} 95--107

\bibitem{feynman:FLP3:1965}
Feynman R~P, Leighton R~B and Sands M 1965 {\em The Feynman Lectures on
  Physics\/} vol~3 (Reading, MA: Addison-Wesley)

\bibitem{sanz:JPCM:2002}
Sanz A~S, Borondo F and Miret-Art\'es S 2002 {\em J. Phys.: Condens. Matter\/}
  {\bf 14} 6109--6145

\bibitem{kocsis:Science:2011}
Kocsis S, Braverman B, Ravets S, Stevens M~J, Mirin R~P, Shalm L~K and
  Steinberg A~M 2011 {\em Science\/} {\bf 332} 1170--1173

\bibitem{aharonov:PRL:1988}
Aharonov Y, Albert D~Z and Vaidman L 1988 {\em Phys. Rev. Lett.\/} {\bf 60}
  1351--1354

\bibitem{prosser:ijtp:1976-1}
Prosser R~D 1976 {\em Int. J. Theor. Phys.\/} {\bf 15} 169--180

\bibitem{sanz:PhysScrPhoton:2009}
Davidovi\'c M, Sanz A~S, Arsenovi\'c D, Bo\v{z}i\'c M and Miret-Art\'es S 2009
  {\em Phys. Scr.\/} {\bf T135} 014009(1--5)

\bibitem{sanz:AnnPhysPhoton:2010}
Sanz A~S, Davidovi\'c M, Bo\v{z}i\'c M and Miret-Art\'es S 2010 {\em Ann.
  Phys.\/} {\bf 325} 763--784

\bibitem{sanz:JRLR:2010}
Bo\v{z}i\'c M, Davidovi\'c M, Dimitrova T~L, Miret-Art\'es S, Sanz A~S and Weis
  A 2010 {\em J. Russ. Laser Res.\/} {\bf 31} 117--128

\bibitem{sanz:EPN:2013}
Davidovi\'c M and Sanz A~S 2013 {\em Europhys. News\/} {\bf 44} 33--36

\bibitem{heisenberg:ZPhys:1925}
Heisenberg W 1925 {\em Z. Phys.\/} {\bf 33} 879--880

\bibitem{rosen:AJP-1:1964}
Rosen N 1964 {\em Am. J. Phys.\/} {\bf 32} 377--379

\bibitem{rosen:AJP-2:1964}
Rosen N 1964 {\em Am. J. Phys.\/} {\bf 32} 597--600

\bibitem{rosen:AJP:1965}
Rosen N 1965 {\em Am. J. Phys.\/} {\bf 33} 146--150

\bibitem{richardson:PRA:2014}
Richardson C~D, Schlagheck P, Martin J, Vandewalle N and Bastin T 2014 {\em
  Phys. Rev. A\/} {\bf 89} 032118

\bibitem{chou:AOP:2016}
Chou C~C 2016 {\em Ann. Phys.\/} {\bf 371} 437--459

\bibitem{rosen:FoundPhys:1986}
Rosen N 1986 {\em Found. Phys.\/} {\bf 16} 687--700

\bibitem{berry:LesHouches:1991}
Berry M~V 1996 {\em Chaos and Quantum Physics\/} Les Houches Lecture Series LII
  (1989) ed Giannoni M~J, Voros A and Zinn-Justin J (Amsterdam: North-Holland)
  pp 251--304

\bibitem{keith:PRL:1991}
Keith D~W, Ekstrom C~R, Turchette Q~A and Pritchard D~E 1991 {\em Phys. Rev.
  Lett.\/} {\bf 66} 2693--2696

\bibitem{cronin:RMP:2009}
Cronin A~D, Schmiedmayer J and Pritchard D~E 2009 {\em Rev. Mod. Phys.\/} {\bf
  81} 1051--1129

\bibitem{wheeler:1978}
Wheeler J~A 1978 {\em The past and the delayed-choice double-slit experiment\/}
  Mathematical Foundations of Quantum Theory (New York: Academic Press)

\bibitem{sanz:foundphys:2015}
Sanz A~S 2015 {\em Found. Phys.\/} {\bf 45} 1153--1165

\bibitem{sanz:JCP-Talbot:2007}
Sanz A~S and Miret-Art\'es S 2007 {\em J. Chem. Phys.\/} {\bf 126}
  234106(1--11)

\bibitem{sanz:AJP:2012}
Sanz A~S and Miret-Art\'es S 2012 {\em Am. J. Phys.\/} {\bf 80}
525--533

\bibitem{talbot:PhilosMag:1836}
Talbot H~F 1836 {\em Philos. Mag.\/} {\bf 9} 401--407

\bibitem{pritchard:PRA:1995}
Chapman M~S, Ekstrom C~R, Hammond T~D, Schmiedmayer J, Tannian B~R, Wehinger S
  and Pritchard D~E 1995 {\em Phys. Rev. A\/} {\bf 51} R14--R17

\bibitem{arndt:OptExpress:2009}
Case W~B, Tomandl M, Deachapunya S and Arndt M 2009 {\em Opt. Express\/} {\bf
  17} 20966--20974

\bibitem{cronin:NJP:2009}
McMorran B~J and Cronin A~D 2009 {\em New J. Phys.\/} {\bf 11} 033021(1--7)

\bibitem{sanz:prb:2000}
Sanz A~S, Borondo F and Miret-Art\'es S 2000 {\em Phys. Rev. B\/} {\bf 61}
  7743--7751

\bibitem{sanz:EPL:2001}
Sanz A~S, Borondo F and Miret-Art\'es S 2001 {\em Europhys. Lett.\/} {\bf 55}
  303--309

\bibitem{sanz:JPA:2005}
Sanz A~S 2005 {\em J. Phys. A: Math. Theor.\/} {\bf 38} 6037--6049

\bibitem{berry:JPA:1996}
Berry M~V 1996 {\em J. Phys. A: Math. Theor.\/} {\bf 29} 6617--6629

\bibitem{sanz:AOP-arxiv:2017}
Tounli J, Alvarado A and Sanz A~S (submitted, 2018)

\end{thebibliography}


\providecommand{\newblock}{}

\end{document}